\documentclass[reprint,aip,jcp,superscriptaddress]{revtex4-1}
\usepackage{graphicx}
\usepackage{multirow}
\usepackage{amsmath}
\usepackage{amssymb}
\usepackage{mathtools}
\usepackage{blindtext}
\usepackage[colorlinks,citecolor=blue,urlcolor=black]{hyperref}
\usepackage{setspace}

\def\ben{\begin{equation}}
\def\een{\end{equation}}

\begin{document}
\title{Density matrix expansion based semi-local exchange hole applied to range separated
density functional theory}

\author{Bikash Patra}
\email{bikash.patra@niser.ac.in}
\author{Subrata Jana}
\email{subrata.jana@niser.ac.in}
\author{Prasanjit Samal}
\email{psamal@niser.ac.in}
%\author{author4}
%\email{author4@niser.ac.in}
\affiliation{School of Physical Sciences, National Institute of Science Education and Research,
HBNI, Bhubaneswar 752050, India.}

\date{\today}

\begin{abstract}
Exchange hole is the principle constituent in density functional theory, which can be used to accurately design
exchange energy functional and range separated hybrid functionals coupled with some appropriate correlation.
Recently, density matrix expansion (DME) based semi-local exchange hole proposed by Tao-Mo gained attention due to 
its fulfillment of some exact constraints. We propose a new long-range corrected (LC) scheme that combines 
meta-generalized gradient approximation (meta-GGA) exchange functionals designed from DME exchange hole coupled 
with the ab-initio Hartree-Fock (HF) exchange integral by separating the Coulomb interaction operator using 
standard error function. Associate with Lee-Yang-Parr (LYP) correlation functional, assessment and bench-marking 
of our functional using well-known test set shows that it performs remarkably well for a broad range of molecular 
properties, such as thermochemistry, noncovalent interaction and barrier height of chemical reactions.
\end{abstract}

\maketitle

\section{Introduction}
Since its advent, the Hohenberg-Kohn-Sham~\cite{ks} density functional theory (DFT) has become one of the most 
sophisticated and widely used computational tools in studying the electronic structure of atoms, molecules and 
solids for its reliable accuracy and computational affordability. Although the formulation is exact, the only 
unknown part of the total energy is exchange-correlation (XC) functional, which needed to be approximate. Designing 
accurately the XC functional is an active and intriguing research field with several new perspectives. Several 
approximations of the XC functional has been reported for past couple of decades with continuous increasing 
accuracy~\cite{PW86,B88,LYP88,BR89,B3PW91,B3LYP,PBE96,VSXC98,HCTH,PBE0,HSE03,AE05,MO6L,TPSS03,revTPSS,PBEsol,SCAN15,
Kaup14,Tao-Mo16}. These proposed density functionals are distinguished through the Jacob's Ladder\cite{jacob's ladder} 
of density functional approximations according to their accuracy and ingredients they used. On the lowest rung of 
the Jacob's ladder, there are local density approximations (LDA)~\cite{VWN80,PW92}, which uses electron density only. 
One rung higher than the LDA, the generalized gradient approximation (GGAs)~\cite{PBE96,PW91,ZWu06} are designed using 
reduced density gradient as its ingredients. Next, meta-generalized gradient approximation (meta-GGAs)~\cite{TPSS03,
MO6L,SCAN15,Tao-Mo16} are proposed using the Kohn-Sham kinetic energy density (KE-KS) along with the reduced density 
gradient. All these semi-local functionals are very successful in describing atomization energies~\cite{dataset,VNS03,
PHao13,LGoerigk10,LGoerigk11}, equilibrium lattice constant~\cite{dataset,YMo16}, equilibrium bond lengths~\cite{dataset,
CAdamo10,YMo16}, surface properties~\cite{YMo16}, cohesive energies~\cite{VNS04} etc.

Even if these functionals enjoy early success, but fail badly to explain Rydberg excitation energy\cite{rydberg}, charge 
transfer excitation\cite{chargetrans}, reaction barrier height and oscillator strength. Missing non-locality~\cite{PSTS08} 
and absence of many electron self-interaction errors (MESI)~\cite{selfinteraction} are the two major drawbacks of semi-local
formalism. Another serious difficulty is that XC potential of the corresponding semi-local functional shows incorrect 
asymptotic behavior and decays faster than $-1/r$\cite{assymp}, where r is the distance of the electron from the nuclei. 
Non-locality of exchange functional, which is missing in semi-local functional is essential~\cite{PSTS08} to describe 
long-range charge transfer, barrier height and dissociation limit of a molecule. Though, the most popular hybrid B3LYP and 
others ~\cite{B3LYP,VNS03} resolve these problems to some extent but it is far from accuracy in several cases. Especially, 
in describing dissociation limit, barrier height, phenomena related to the fractional occupation number, dramatic failures 
of the hybrid functionals are observed. Not only that, in hybrid functional, the XC potential decays like $-c/r$, where c is 
the fraction of mixed Hartree-Fock exchange. One of the interesting findings is to mix Hartree-Fock (HF) with that of 
semi-local functional by using range separation of coulomb interaction operator (may be long range or short range). This 
range separation technique~\cite{HSE03,camb3lyp,lcwpbe,lcwpbeh,bnl} also enable us to change the range of semi-local and HF 
part. Using this method one can mix semi-local form of DFT in long range or short range depending on its requirement. It 
has been shown that for molecules the long range corrected HF is more effective~\cite{camb3lyp,lcwpbe,lcwpbeh,bnl} and for 
solid state calculation short range HF exchange~\cite{HSE03} is superior from computational point of view. All these ideas 
are used to design the long range corrected  LC-$\omega$PBE~\cite{lcwpbe}, LC-BLYP\cite{camb3lyp}, which reproduces 
intriguing result and establish superiority over B3LYP in many cases and HSE06~\cite{HSE03}, which uses short range HF and 
mainly applied for the solid state system. In some functionals like CAM-B3LYP\cite{camb3lyp}, LC-$\omega$PBEh\cite{lcwpbeh} both 
the HF exchange and DFT counterpart are incorporated over the whole range by using generalize parametrization of coulomb 
interaction operator.

The range separated functionals are designed using the semi-local exchange hole constructed from spherical averaged exchange 
hole~\cite{PW86,BR89,Tao-Mo16,Taobook10} or using reverse engineering technique~\cite{EP98,tpsshole,Lucian13}. The CAM-B3LYP 
functional is designed using LDA exchange hole by passing in-homogeneity through Thomas-Fermi wave-vector~\cite{gga}. 
Later, LC-$\omega$PBE, LC-$\omega$PBEh and HSE06 are designed using PBE exchange hole. Though spherical averaged exchange hole 
is not available for PBE exchange, the way it is designed using system averaged exchange hole. The present aim of this paper 
is to design a long range corrected meta-GGA level exchange  energy functional using recently developed density matrix expansion
(DME) based exchange hole. The density matrix expansion based exchange hole has unique properties like correct uniform density 
limit, correct small u expansion etc., which  were previously lack in the exchange hole expansion proposed by Becke and Scuseria 
and its coworker. The most successful meta GGA exchange hole proposed by Tao-Mo using DME satisfy the above mentioned properties 
with addition its large u limit also converges. With all these exact criteria it is always interesting to design a range separated 
functional using this exchange hole. The present attempt of this paper is in that direction. We use the semi-local exchange hole 
in short range limit and long range is fixed with HF exchange. The exchange energy proposed here is tested with LYP~\cite{LYP88} 
correlation. The present paper is organized as follows. In the next section, we briefly describe the theoretical background of 
range separated functional and next our methods for designing the long range corrected meta-GGA level range separated functional. 
Next section our functional is bench-marked with several molecular test set. For comparison with our functional, we consider 
CAM-B3LYP, HSE06, LC-PBEh, LC-$\omega$PBE functionals because they are the commonly used functionals. 

\section{Theoretical Background}
\begin{figure}[h]
\includegraphics[width=.80\linewidth]{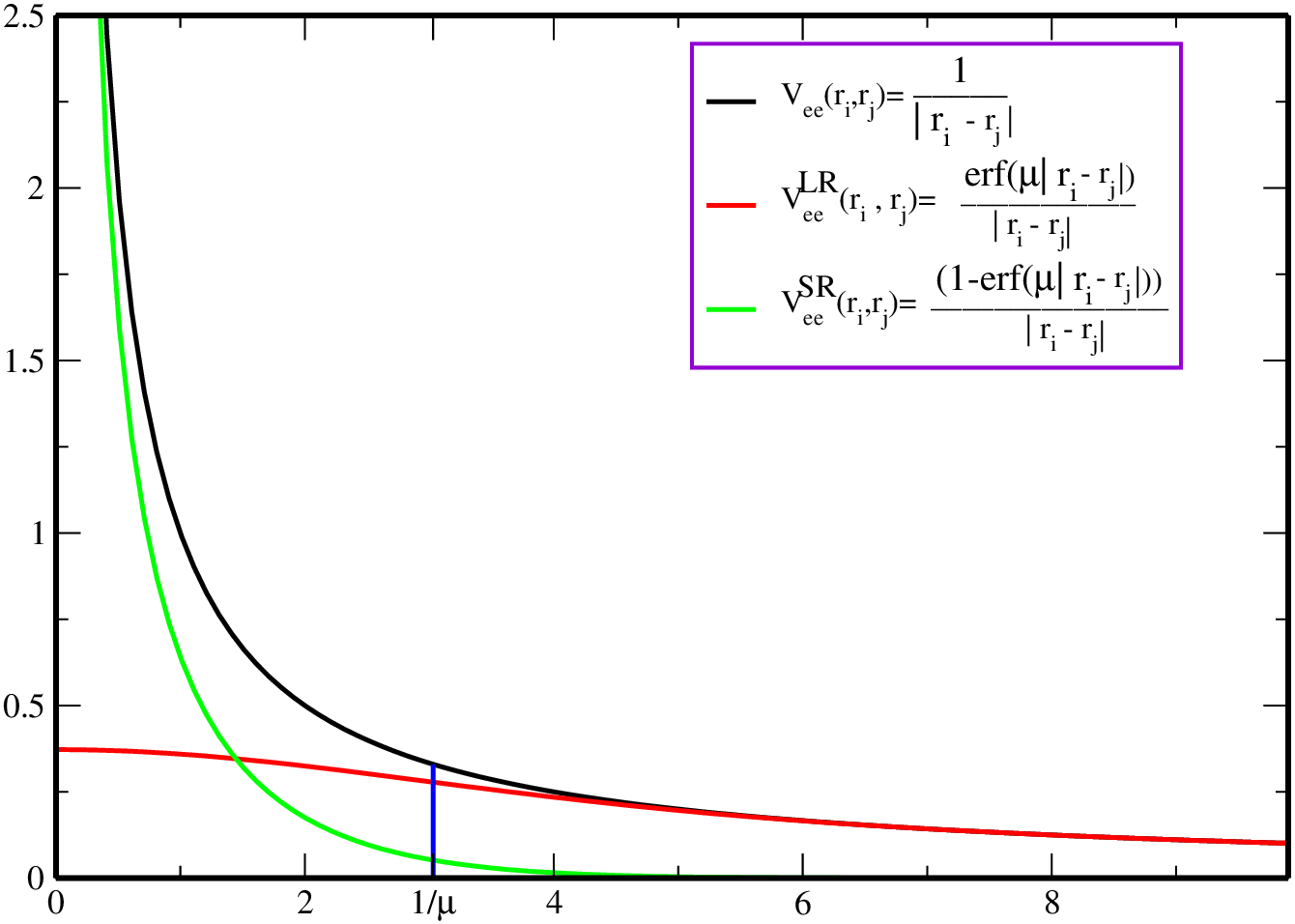}\hfill
\caption{Full-range, long range and short range part of coulomb interaction for range separation parameter $\mu=0.33$ versus
electron-electron distance $|\boldsymbol{r}_i-\boldsymbol{r}_j|$.}
\label{fig1}
\end{figure}

 Let  the Hamiltonian for N electron system,
 \begin{equation}
 \mathcal{H}=\sum_{i} h(i)+\sum_{i>j} \frac{1}{r_{ij}},
 \label{eq1}
  \end{equation}
where $h$ represents the one electron part of the Hamiltonian which includes kinetic energy of the electrons and electron-nuclei 
interaction and  $r_{ij}=|\boldsymbol{r}_i-\boldsymbol{r}_j|$. Now in range separated functionals the electron-electron coulomb 
interaction $ V_{ee}({\boldsymbol{r}_i}, {\boldsymbol{r}_j})=\frac{1}{|\boldsymbol{r}_i-\boldsymbol{r}_j|}$  between an electron 
at  $\boldsymbol{r}_i$ and another at $\boldsymbol{r}_j$ can be separated into a long range (\textbf{LR}) and a short range part 
(\textbf{SR})~\cite{savin97,savin02} as,
\begin{equation}
  \frac{1}{r_{ij}}=\underbrace{\frac{1-g(r_{ij})}{r_{ij}}}_{SR}+\underbrace{\frac{g(r_{ij})}{r_{ij}}}_{LR}.
  \label{eq2}
\end{equation}   
From various possible choices of the function $g$\cite{LRSRS1,LRSRS2,LRSRS3}, the most suitable from both physical and computational 
point of view is, $g=erf(\mu r_{ij})$, where $\mu$ is a parameter. For this choice, the first term of Eq.(\ref{eq2}) approaches to 
$0$  as $r_{ij}\rightarrow \infty$ and second  term goes to $2\mu/\sqrt{\pi}$ when $r_{ij}\rightarrow 0$. The parameter $\mu$ can  
be treated as a cutoff between SR and LR part as shown in Fig.(\ref{fig1}).

In the range separated Kohn Sham scheme the ground state energy of an electronic system is expressed as\cite{GSE1,GSE2} 
\begin{equation}
E= m\underset\psi in \lbrace<\psi|\hat{T}+\hat{V}_{ne}+\hat{V}_{ee}^{LR}|\psi>+E_{Hxc}^{SR}[\rho_\psi]\rbrace
\label{eq3}
\end{equation}
where $\hat{T}$ is the kinetic energy operator, $\hat{V}_{ne}$ is the electron-nuclei interaction operator, $\hat{V}_{ee}^{LR}$
is the operator for long range part of electron-electron interaction, $E_{Hxc}^{SR}$ is an energy functional, which include short
range Hartree, exchange and correlation energy, and $\psi$ is a multi-determinant wave function. The minimizing wave function 
$\psi_{min}^{LR}$ simultaneously minimizes the long range interacting effective Hamiltonian and give the exact density of the 
interacting many body system. For $\mu=0$ as the  long range interaction vanishes, we get back the standard Kohn-Sham system, 
and on the other hand for $\mu=\infty$ the wave function  based formulation of electronic structure is recovered. Up to this, 
the above theory is exact. Now if we replace the multi-determinant  wave function in Eq.(\ref{eq3}) by a single Slater determinant 
wave function $\phi$ then the ground state energy of range separated hybrid scheme \cite{RSH} becomes
\begin{equation}
\begin{split}
E_{RSH}^0= <\phi_0|\hat{T}+\hat{V}_{ne}|\phi_0>+E_H[\rho_0] \\ +E_{x,HF}^{LR}[\phi_0] +E_{xc}^{SR}[\rho_0],
\end{split}
\label{eq4}
\end{equation}
where $\phi_0$ is the Slater determinant which minimizes energy functional and $\rho_0$ is the associated density.

If the range separation is applied only on exchange part then it is called long range corrected functional (LC) and the above 
expression can be written as, 
\begin{equation}
\begin{split}
E_{RSH}^0= <\phi_0|\hat{T}+\hat{V}_{ne}|\phi_0>+E_H[\rho_0]+E_{x,HF}^{LR}[\phi_0]\\  +E_{x}^{SR}[\rho_0]+E_c[\rho_0]
\end{split}
\label{eq5}
\end{equation}
In the above expression the long range part of the exchange interaction can be written as,
\begin{equation}
\begin{split}
E_{x,HF}^{LR}[\phi_0]=-\frac{1}{2}\sum_{\sigma}\sum_{i,j}^{occ}\int\int \phi_{i\sigma}^*({\bf{r}}_i)\phi_{j\sigma}^*({\bf{r}}_j)
\frac{erf(\mu r_{ij})}{r_{ij}}\\ \phi_{j\sigma}({\bf{r}}_i)\phi_{i\sigma}({\bf{r}}_j)d{\bf{r}_i}d{\bf{r}_j}
\end{split}
\label{eq6}
\end{equation}
where $\phi_{i\sigma}$ is the ith $\sigma$-spin molecular orbital. The short range exchange functional is defined by,
\begin{equation}
\begin{split}
E_x^{SR}[\rho]=\frac{1}{2}\int\int\frac{\rho({\bf{r}_i})(1-erf(\mu r_{ij}))\rho_x({\bf{r}_i},{\bf{r}_j})}{r_{ij}}\\
d{\bf{r}_i}d{\bf{r}_j}
\end{split}
\label{eq7}
\end{equation}
where $\rho_x$ is the exchange hole and conventionally defined as $\rho_x({\bf{r}_i},{\bf{r}_j})=-|\rho_1({\bf{r}_i},{\bf{r}_j})|^2/
2\rho({\bf{r_i}})$, with $\rho_1({\bf{r}_i},{\bf{r}_j})$ is the first-order reduced density matrix. Spin polarized form of the equation 
(\ref{eq7}) can be written using spin scaling\cite{spin scaling} relationship
\begin{eqnarray}
 E_x^{SR}[\rho\uparrow,\rho\downarrow]=(E_x^{SR}[2\rho\uparrow]+E_x^{SR}[2\rho\downarrow])/2
\end{eqnarray}

\section{Present Long Range Corrected Hybrid Functional}
Now, we will propose range separated hybrid functional based on recently developed DME exchange hole of Tao-Mo\cite{Tao-Mo16}. The 
DME exchange hole of Tao-Mo has following semi-local form
\begin{equation}
 \begin{split}
\rho_x({\bf{r}},{\bf{u}})=-\frac{9\rho}{2}\frac{j_1^2(k u)}{k^2 u^2}-\frac{105 j_1(k u)j_3(ku)}{ k^4 u^2}\mathcal{G}\\ 
-\frac{3675 j_3^2(k u)}{8 k^6 u^4}\mathcal{H}
\end{split}
\label{eq8}
\end{equation}
where $\mathcal{G}=3(\lambda^2-\lambda+\frac{1}{2})(\tau-\tau^{unif}-\frac{|\nabla\rho|^2}{72\rho})-(\tau-\tau^{unif})+7(2\lambda-1)^2
\frac{|\nabla\rho|^2}{18\rho}$ and $\mathcal{H}=(2\lambda-1)^2\frac{|\nabla\rho|^2}{\rho} $. Tao-Mo proposed exchange hole uniquely 
recovers (i) correct uniform density limit, (ii) correct small $u$ expansion proposed be Becke~\cite{BR89}, and (iii) its large $u$ 
limit also converges. To take care the in-homogeneity of the system the parameter $k$ is set to be $k=fk_F$, where $k_F$ is the 
Fermi momentum for homogeneous electron gas and $f$ is fixed from sum rule of exchange hole by extrapolating its low gradient 
limit and large gradient limit. The obtained form is $f=[1+10(70y/27)+\beta y^2]^{1/10}$, where $y=(2\lambda-1)^2p$, $p=|\nabla\rho|^2/
(2k_F\rho)^2$ and $\beta$ is a parameter. We take the value of $\lambda=0.6866$ and $\beta=79.873$, as prescribed in the Tao-Mo 
paper~\cite{Tao-Mo16}. Using this exchange hole in equation (\ref{eq7}) we get
\begin{eqnarray}
E_x^{SR}[\rho]=\frac{1}{2}\int\rho(\mathbf{r})(\mathcal{M}+\mathcal{N}+\mathcal{Q}) d^3r 
\label{eq9}
\end{eqnarray}
 where,
 \begin{widetext}
\begin{eqnarray}
\mathcal{M}=-\frac{9\pi\rho}{2k^2}\Big[1-\frac{8}{3}a\Big\{\sqrt{\pi}erf(\frac{1}{2a})+
(2a-4a^3) exp(-\frac{1}{4a^2})-3a+4a^3\Big\}\Big]
\label{eq10}
\end{eqnarray}
\begin{eqnarray}
 \mathcal{N}=-\frac{35\pi}{3k^4}\mathcal{G}\Big[1+24a^2\Big\{(20a^2-64a^4)exp(-\frac{1}{4a^2})
 -3-36a^2+64a^4+10\sqrt{\pi}erf(\frac{1}{2a})\Big\}\Big]
\label{eq11}
\end{eqnarray}
\begin{eqnarray}
 \mathcal{Q}=-\frac{245\pi}{48k^4}\mathcal{H}\Big[1+\frac{8}{7}a\Big\{(-8a+256a^3-
 576a^5+3849a^7-122880a^9)exp(-\frac{1}{4a^2})+\nonumber\\24a^3(-35+224a^2-1440a^4+5120a^6)+2\sqrt{\pi}(-2+60a^2)erf(\frac{1}{2a})
 \Big\}\Big],
\label{eq12}
\end{eqnarray}
\end{widetext}
%\begin{eqnarray}
 
%\end{eqnarray}
with $a = \mu/2k$. Most of the previous implementations of LC functional uses global system independent screening parameter. However 
there are studies which uses position dependent screening parameter\cite{pos-screen}, but those are nontrivial to implement. For 
the parameter $\mu$, same value is used as in CAM-B3LYP functional, $\mu = 0.33$. For the correlation functional, we use the one 
electron self interaction free Lee-Yang-Parr (LYP) \cite{lyp} correlation functional.

\section{Computational Details}
We have implemented Eq. (\ref{eq9}) by locally modifying the NWChem-6.6 \cite{nwchem} program. We use the same value of $\mu=0.33$
in all parts of the code. Medium grid is used for the evaluation of the exchange-correlation contribution to the density functional.
All the results of benchmark calculations reported in this article are obtained self-consistently using NWChem program.  Spin 
unrestricted calculation have been done for open shell system. Deviations are defined as theory - experiment and reported as
mean errors (ME) and mean absolute errors (MAE) and in some cases maximum deviation from standard value. For all the cases under 
study we compare the results of our range separated functional (DME-RS) with four others popularly known range separated functional 
CAM-B3LYP \cite{camb3lyp}, HSE06 \cite{HSE03}, LC-$\omega$PBE \cite{lcwpbe} and LC-$\omega$PBEh \cite{lcwpbeh}. 
\section{Results} 
For bench-marking our functional the following databases have been used- i) AE17 for atomic energy, ii) G2/97 for atomization energy, 
iii) IP13/03 for ionization potential, iv) EA13/03 for electron affinity, v) PA8/06 for proton affinity,  vi) NCCE31/05 for noncovalent 
interaction,vii) HTBH38/08 for hydrogen transfer barrier height, viii) NHTBH38/08 for non-hydrogen transfer barrier height,
complexes, ix) $\pi$TC13 for the thermochemistry of $\pi$ system, x) ABDE12 for alkyl bond dissociation energy, x) HC7/11 for 
Hydrocarbon chemistry, xi) ISOL6/11 for Isomerization energies of large molecules, xii) DC7/11 for Difficult cases.
%%%%%%%%%%%%%%%%%%%%%%%%%%%%%%%%%%%%%%%%%%%%%%%%%%%%%%%%%%%%%%%%%%%%%%%%%%%%%%%%%%%%%%%%%%%%%%%%%%%%%%%%%%%%%%%%%%%%%%%%%%%%%
\subsection{Atomic energy}
For the functionals under study, we have calculated the total energies of atoms from H to Cl (AE17)\cite{dataset}. For all atoms, we 
have used the 6-311++G(3df,3pd) basis set except for Helium, for which aug-cc-pVQZ basis set is used. We compared the calculated total 
energies with the accurate non-relativistic values\cite{atomicenergy}. CAM-B3LYP has the smaller MAE among the tested functionals. DME-RS 
gives comparable results with CAM-B3LYP and LC-$\omega$PBEh has the largest MAE in this case.
\begin{table}[h!]
\caption{Summary  of deviations of calculated total energy of atoms from reference values. All values are in eV.}
\begin{tabular}{l  c  c c c c c c c c c }
\hline\hline
 Functional & ME  & MAE & Max(+) & Min(-) \\
 \hline
 DME-RS                  & -0.341  & 0.382 & 0.887 (Al) & -0.217 (Be) \\
 CAM-B3LYP               & -0.140  &  0.332 & 0.615 (Al) & -0.487 (Be) \\
 HSE06                   & 1.322   & 1.334  & 0.099 (H) & -3.172 (Cl)\\
 LC-$\omega$PBEh         & 1.735  & 1.745  & 0.078 (H) & -3.871 (Cl) \\
 LC-$\omega$PBE          & 1.564  & 1.584  &0.169 (H)&-3.837 (Cl) \\
 \hline\hline
\label{atmic energy}
\end{tabular}
\end{table}
%%%%%%%%%%%%%%%%%%%%%%%%%%%%%%%%%%%%%%%%%%%%%%%%%%%%%%%%%%%%%%%%%%%%%%%%%%%%%%%%%%%%%%%%%%%%%%%%%%%%%%%%%%%%%%%%%%%%%%%%%%%%%%  
      
\subsection{Thermochemistry}
We have calculated atomization energies (AE) of 148 molecules for the G2/97 test set using MP2(full)/6-31G* optimized
geometry.\cite{g2geometry}The atomization energy of  molecule is defined as the energy difference between the 
total energy of a molecule and the sum of energy of its constituent free atoms, all at 0 K. We have compared our 
calculated atomization energy with the result of CCSD(T) \cite{ccsdt}, which is gold-standard in quantum chemistry. 
Table (\ref{atomization energy}) shows, CAM-B3LYP gives the smallest error in AEs, followed by DME-RS. LC-$\omega$PBEh 
is the least accurate for AEs among the tested functionals.

Ionization potential (IP) and electron affinity (EA) is the amount of total energy difference between the ion and 
corresponding neutral atom or molecule, all at 0 K. We have calculated IP for IP13 \cite{ip13} database and EA for 
EA13/3\cite{ea13} data base and QCISD/MG3\cite{minnesotadatabase} level optimized geometry have been taken
for both cases. For IP, HSE06 performs best with smallest MAE of 0.139 eV. For both IP and EA, DME-RS is comparable
in accuracy with the other range separated functionals.

Proton affinity (PA) is the amount of energy released when a proton is added to a species at its ground state. We have 
calculated PA of PA8 \cite{pa1,pa2} data-set with MP2/6-31G(2df,p) level optimized geometry.  6-311++G(3df,3pd) basis 
set is used for all calculation. From Table (\ref{proton affinity}) DME-RS gives the smallest mean absolute error which 
is 0.051 eV. All the investigating functionals overestimate PA except for CAM-B3LYP.
\begin{table}[h!]
\caption{Summary of deviations of the calculated atomization energy from reference value. All quantities are in kcal/mol.}
\begin{tabular}{l  c  c  c c c c c c c c c }
\hline\hline
 Functional &  ME & MAE & Max (+) & Max(-)  \\
 \hline
DME-RS             & -4.494 & 5.099 & 8.253 (BeH) & -25.8766 (C$_2$Cl$_4$) \\
CAM-B3LYP          & 1.199 &  4.329 & 15.748 (C$_2$NH$_7$)  & -20.276 (SiCl$_4$) \\
HSE06              &  -4.610 & 5.195  & 6.133 (C$_5$H$_5$N) & -35.505 (SiF$_4$)\\
LC-$\omega$PBEh    &  2.276 & 5.735 & 22.317 (C$_5$H$_5$N)  & -24.908 (SiF$_4$)\\
LC-$\omega$PBE     &   1.746 & 5.362 &   21.765 (C$_5$H$_8$) & -18.140 (P$_2$)\\
\hline\hline

\label{atomization energy}
\end{tabular}
\end{table}

\begin{table}[h!]
\caption{Summary of deviations of the calculated Ionization potential from reference value. All quantities are in eV.}
\begin{tabular}{l  c  c  c c c c c c c c c }
\hline\hline
Functional & ME  & MAE & Max (+) & Min (-)\\
 \hline
 DME-RS            & 0.083 & 0.169  & 0.426 (O) & -0.223 (P) \\
 CAM-B3LYP         &  0.158 &0.196  &  0.499 (O)& -0.149 (P) \\
 HSE06             & 0.108  &0.139  & 0.333 (O$_2$) & -0.173 (Cl$_2$)\\
 LC-$\omega$PBEh   & 0.147  &0.167  & 0.326 (O$_2$) & -0.126 (Cl$_2$) \\
 LC-$\omega$PBE    & 0.239  & 0.239 & 0.502 (O)     &  -\\
 \hline\hline
\label{ionization potential}
\end{tabular}
\end{table}

\begin{table}[h!]
\caption{Deviation of the calculated electron affinity from reference value. All quantities are in eV.}
\begin{tabular}{l  c  c  c c c c c c c c c }
\hline\hline
Functional & ME  & MAE & Max (+) & Min (-)\\
 \hline
 DME-RS            & -0.077  & 0.128 &  0.251 (Cl$_2$) & -0.238 (Si) \\
 CAM-B3LYP         & 0.022   & 0.084 & 0.304 (Cl$_2$)  & -0.141 (Si) \\
 HSE06             & -0.068  & 0.123 & 0.186 (Cl$_2$) & -0.307 (OH)\\
 LC-$\omega$PBEh   & -0.052  & 0.105 &  0.166 (C)    & -0.242 (OH)\\
 LC-$\omega$PBE    & -0.027  &  0.089&  0.180 (C)    & -0.165 (PH$_2$)\\
 \hline\hline
\label{electron affinity}
\end{tabular}
\end{table}

\begin{table}[h!]
\caption{Summary of deviation of the calculated proton affinity from reference value. All quantities are in eV.}
\begin{tabular}{l  c  c  c c c c c c c c c }
\hline\hline
Functional & ME  & MAE & Max (+) & Min (-)\\
 \hline
 DME-RS          &  0.012  & 0.051 &  0.099 (PH$_3$)    & -0.084 (H$_2$) \\
 CAM-B3LYP       & -0.047  & 0.084 & 0.048 (C$_2$H$_2$) & -0.117 (H$_2$) \\
 HSE06           &  0.077  &  0.077& 0.211 (C$_2$H$_2$) &  -\\
 LC-$\omega$PBEh & 0.470   & 0.470 &0.642 (PH$_3$) & -\\
 LC-$\omega$PBE  & 0.086   & 0.086 &0.243 (C$_2$H$_2$)& - \\
 \hline\hline
\label{proton affinity}
\end{tabular}
\end{table}

\subsection{Binding energy of weakly interacting system}
Weakly interacting systems remain a challenging class of materials to describe accurately  within  the  DFT  approaches  in  
practice because of the long range nature of the weak interactions such as Van Der Waals interaction, which is absent in 
semi-local functionals. However, introducing kinetic energy in the density functional can substantially reduce the error 
of LSDA and GGA functional for predicting the binding energy of weakly interacting system\cite{dataset}. We have tested 
our meta-GGA range separated functional for NCCE31/05 database. We have taken the equilibrium geometries for single point 
calculation from Ref. 21. 6-311++G (3df,3pd) basis set is used for all the molecules except for the inert gas related 
molecules for which we have used aug-cc-pVQZ. The summary of the performances of the tested functionals are listed in 
Table (\ref{ncce}). All the investigating LC functionals performing very well. From Table (\ref{ncce}) we can say DME-RS is 
the best choice in this case.

\begin{table}[h!]
\caption{Mean Absolute error (MAE) for the noncovalent complexation energies database
(NCCE31/05) and its secondary databases for the functionals shown in each row. All values are in kcal/mol.}
\label{ncce}
\begin{tabular}{lccccccccccc}
\hline\hline
&DME-RS & CAM-B3LYP & HSE06 & LC-$\omega$PBEh & LC-$\omega$PBE \\
 
\hline
 ME & -0.27 & -0.24& -0.46 & -0.45 & -0.70 \\
 MAE&  0.45  & 0.58 & 0.71 &  0.56 &  0.78  \\
 
\hline\hline
\end{tabular}
\end{table}

\subsection{Thermochemistry of $\pi$ system}
The molecules with $\pi$ bond are largely dominated by multi configurational state functions than $\sigma$-bonded molecules 
due to their small HOMO-LUMO gap. We are investigating the performance of the range separated functionals to explain the 
properties of $\pi$ system. The database ($\pi$TC13) of $\pi$ system contains three secondary database - i) $\pi$IE3/06- 
isomeric energy differences between allene and propyne and higher homologs. ii) PA-CP5/06- Proton affinities of five conjugated 
polyenes. iii) PA-CP5/06- Proton affinities of five conjugated Schiff bases\cite{pisystem}. We have taken MP2/6-31+G(d,p) level 
optimized geometries from Minessota database~\cite{minnesotadatabase} and 6-311++G(3df,3pd) basis set is used for all the 
geometries. CAM-B3LYP gives the smallest MAE with 3.41 kcal/mol, followed by DME-RS with MAE of 4.19 kcal/mol.

\begin{table}[h!]
\centering
\caption{Mean Absolute error (MAE) for the $\pi$TC13 and its secondary databases for the functionals shown in each row.
All values are in kcal/mol.}
\label{pi}
\begin{tabular}{lccccccccccc}
\hline\hline
Functional &$\pi$TC13  &$\pi$IE3/06 & PA-CP5/06  & PA-SB5/06  \\
\hline
 DME-RS          & 4.19 & 0.99 & 5.19 & 5.12 \\
 CAM-B3LYP       & 3.41 & 2.37 & 3.61 & 3.82 \\
 HSE06           & 6.54 & 4.94 & 6.96 & 7.09 \\
 LC-$\omega$PBEh & 4.79 & 2.39 & 5.45 & 5.58 \\
 LC-$\omega$PBE  & 4.24 & 0.98 & 4.92 & 5.51 \\
  \hline\hline
\end{tabular}
\end{table}

\subsection{Barrier heights of chemical reactions}
Semi-local density functionals very often fail to describe the reaction barrier height. Most of the time
it gives the transition state to a lower energy state than reactants or products, giving rise to
negative barrier height. We can associate this problem of semi-local functionals with self interaction
error, because the transition states have stretched bonds and as a result, SIE\cite{sie} may be large.
We have calculated forward and reverse barrier height of 19 hydrogen transfer reaction from HTBH38/04
data-set and same for 19 non-hydrogen-transfer reaction from HTBH38/04\cite{bh1,bh2} data set. Further,
the set NHTBH38/04 is subdivided into the set of six heavy-atom transfer reactions, eight nucleophilic 
substitution reaction, five association and unimolecular reactions. For all the calculation 6-311++G(3df,3pd) 
basis set is used. Equilibrium geometries and reference values are taken from Minessota database
~\cite{minnesotadatabase}. From table (\ref{rbh}), LC-$\omega$PBE gives the lowest MAE for hydrogen transfer 
reaction, and DME-RS gives the minimumMAE for non-hydrogen transfer reaction. Overall DME-RS and LC-$\omega$PBE
gives comparable result for barrier height.

\begin{table*}%[h!]
\centering
\caption{Deviations from experiment of barrier heights of chemical reactions computed using the 6-311++G (3df,3pd)
basis set. All values are in kcal/mol.}
\label{rbh}
\begin{tabular}{@{}lcccccccccccccc@{}}
\hline\hline
\multicolumn{3}{c}{}         &  & \multicolumn{11}{c}{Non-hydrogen-transfer reactions of the NHTBH38 set}  \\
\cline{5-15}
      & \multicolumn{2}{c}{HTBH38} &  & \multicolumn{2}{c}{Heavy-atom} &  & \multicolumn{2}{c}{nucleophilic} &  & \multicolumn{2}{c}{Unimolecular and} &  & \multicolumn{2}{c}{} \\
      & \multicolumn{2}{c}{Hydrogen transfer (38)} &  & \multicolumn{2}{c}{transfer (12)} &  & \multicolumn{2}{c}{substitution (16)} &  & \multicolumn{2}{c}{association (10)} &  & \multicolumn{2}{c}{Full NHTBH38} \\

\cline{2-3}\cline{5-6}\cline{8-9}\cline{11-12}\cline{14-15}
 Functional & ME  & MAE  &  &    ME & MAE  &  &   ME & MAE  &  &  ME & MAE &  & ME & MAE\\
\hline
 DME-RS             & -0.88  & 2.17  && -2.49  & 2.63 && 0.05  & 1.17 && 0.64 & 2.04 && -0.59 & 1.86\\
 CAM-B3LYP          & -3.00  & 3.29  && -5.47  & 5.47 && -0.88 & 1.19 && -0.54& 1.76 && -2.24 & 2.69\\
 HSE06              & -3.45  & 3.48  && 2.68   & 12.37&& -1.34 & 1.53 && -0.55& 1.83 &&  0.14 & 5.04\\
 LC-$\omega$PBEh    & -3.28  & 3.31  && -4.45  & 4.45 && 0.08  & 0.95 && -0.10& 2.26 && -1.39 & 2.40\\
 LC-$\omega$PBE     & -0.57  & 1.26  && -0.19  & 2.11 && 2.95  & 2.95 && 1.41 & 2.34 && 1.55  & 2.52\\
 \hline\hline
\end{tabular}
\end{table*}
\begin{table*}%[h!]
\caption{Summary of deviations of alkyl bond dissociation energy, hydrocarbon chemistry, isomerization energies and 
difficult cases. All values are in kcal/mol}
\label{my-label}
\begin{tabular}{lcccccccccccccc}
\hline\hline
 & \multicolumn{2}{c}{Alkyl bond} &  & \multicolumn{2}{c}{} &  & \multicolumn{2}{c}{} &  & \multicolumn{2}{c}{} \\
 & \multicolumn{2}{c}{dissociation energy} &  & \multicolumn{2}{c}{Hydrocarbon chemistry} &  & \multicolumn{2}{c}{Isomerization energies} &  & \multicolumn{2}{c}{Difficult cases} \\
 \cline{2-3}\cline{5-6}\cline{8-9}\cline{11-12}
 Functional&  ME & MAE & &  ME  &   MAE  &  &  ME &  MAE &  & ME  &  MAE        \\
 \hline
 DME-RS   & -9.08  & 9.08 & & -5.37  & 5.37  &  & -1.96 & 1.96 &  & -15.69 & 15.69   \\
 CAM-B3LYP& -6.63  & 6.63 & & -5.35  & 5.35  &  & -1.80 & 2.03 &  & -4.40  & 9.53    \\  
 HSE06    & -9.38  & 9.38 &  &  0.14  & 5.92  &  & -1.12 & 1.42 &  & -21.21 & 23.07  \\
 LC-$\omega$PBEh & -6.42 &  6.42 &  &  9.22  & 13.65 &  & -0.65 & 1.67 &  & -1.15  & 17.54 \\
 LC-$\omega$PBE  & -5.52  & 5.52 & &  16.05  & 20.07 &  & -1.05 & 1.57 &  &  4.43  & 15.44   \\
 \hline\hline
\end{tabular}
\end{table*}
\subsection{Alkyl bond dissociation energies, Hydrocarbon chemistry, Isomerization energies of large molecules and Difficult cases}
Alkyl bond dissociation energy database (ABDE12)\cite{minnesotadatabase} contains two subsets ABDE4/05 and ABDEL8\cite{pa2}.
ABDE4 includes four bond dissociation energies of R-X organic molecules, where R = methyl and isopropyl, and X = CH$_3$ and 
OCH$_3$. For subset ABDEL8 contains eight molecules, with R = ethyl and tert-butyl and X = H, CH$_3$, OCH$_3$, OH. B3LYP/6-31G(d) 
level optimized geometries are taken from Minessota database~\cite{minnesotadatabase}. For Hydrocarbon chemistry, isomerization 
energy and difficult cases, we have used respectively HC7\cite{dataset}, ISOL6\cite{dataset} and DC7\cite{dataset} databases. For 
all these cases 6-311++G (3df,3pd) basis set is used. For alkyl bond dissociation energy LC-$\omega$PBE performs best with 
MAE 5.52 kcal/mol. In case of hydrocarbon chemistry DME-RS, CAM-B3LYP and HSE06 has the comparable result, while LC-$\omega$PBE 
and LC-$\omega$PBEh perform worst. For Isomerization energies of large molecules, HSE06 achieves smallest MAE, while other range 
separated functionals also perform well. For difficult cases, CAM-B3LYP has the smallest MAE of 9.53 kcal/mol, followed by 
LC-$\omega$PBE0 and DME-RS. HSE06 has the worst performance for difficult cases.

\subsection{Dissociation energy}
Dissociation energy is the amount of energy needed to break every chemical bond in a molecule by separating all of its' constituent atoms. 
Local or semi-local density functional fail to describe the dissociative nature of symmetrical radical cation like $H_2^+$, $He_2^+$ etc. 
The failure of conventional density functional to explain the dissociative nature of one electron molecule $H_2^+$ is an indication of 
self-interaction error. This is due to the fact that traditional density functionals suffer from delocalization error\cite{delocalization}. 
In figure (\ref{fig:sub1}) we have compared the dissociation curve of $H_2^+$ from range separated functional with the Hartree-Fock result, 
which is exact in this case. Albeit all the functionals give same equilibrium bond length, the error increases as we increase the bond length 
away from equilibrium bond length. LC-$\omega$PBE and DME-RS present almost same result, while all the other functionals deviate too much from HF 
result and HSE06 gives the worst performance.

We have also considered the dissociation nature of NaCl molecule, which is an ionic pair when the inter-atomic separation R is not very 
far from equilibrium distance and because of $IP_{Na}>EA_{Cl}$, it dissociates into neutral Na and Cl atom at infinite inter-molecular separation. 
The critical length\cite{critical length} $R_c$ at which sudden charge transfer occurs is given by $R_c=1/(IP_{Na}-EA_{Cl})$. From experimental
value\cite{experimental value} of $IP_{Na}$ and $EA_{cl}$ we get $R_c\approx 9.4$ \AA. HF underestimated the critical length $R_c=5.72$ {\AA}  due 
to overestimation of $IP_{Na}-EA_{Cl}$ difference. DME-RS and LC-$\omega$PBEh almost overlap and gives $R_c=8.51$ {\AA}. From LC-$\omega$PBE and
CAM-B3LYP we obtain $R_c$ equal to 8.93 {\AA} and 9.7 {\AA} respectively. HSE06 completely fails in this case and calculation does not
converge for inter-atomic separation greater than 9.4 {\AA}.

\begin{figure}%[!tbp]
  \centering
  \begin{minipage}[b]{0.4\textwidth}
    \includegraphics[width=\textwidth]{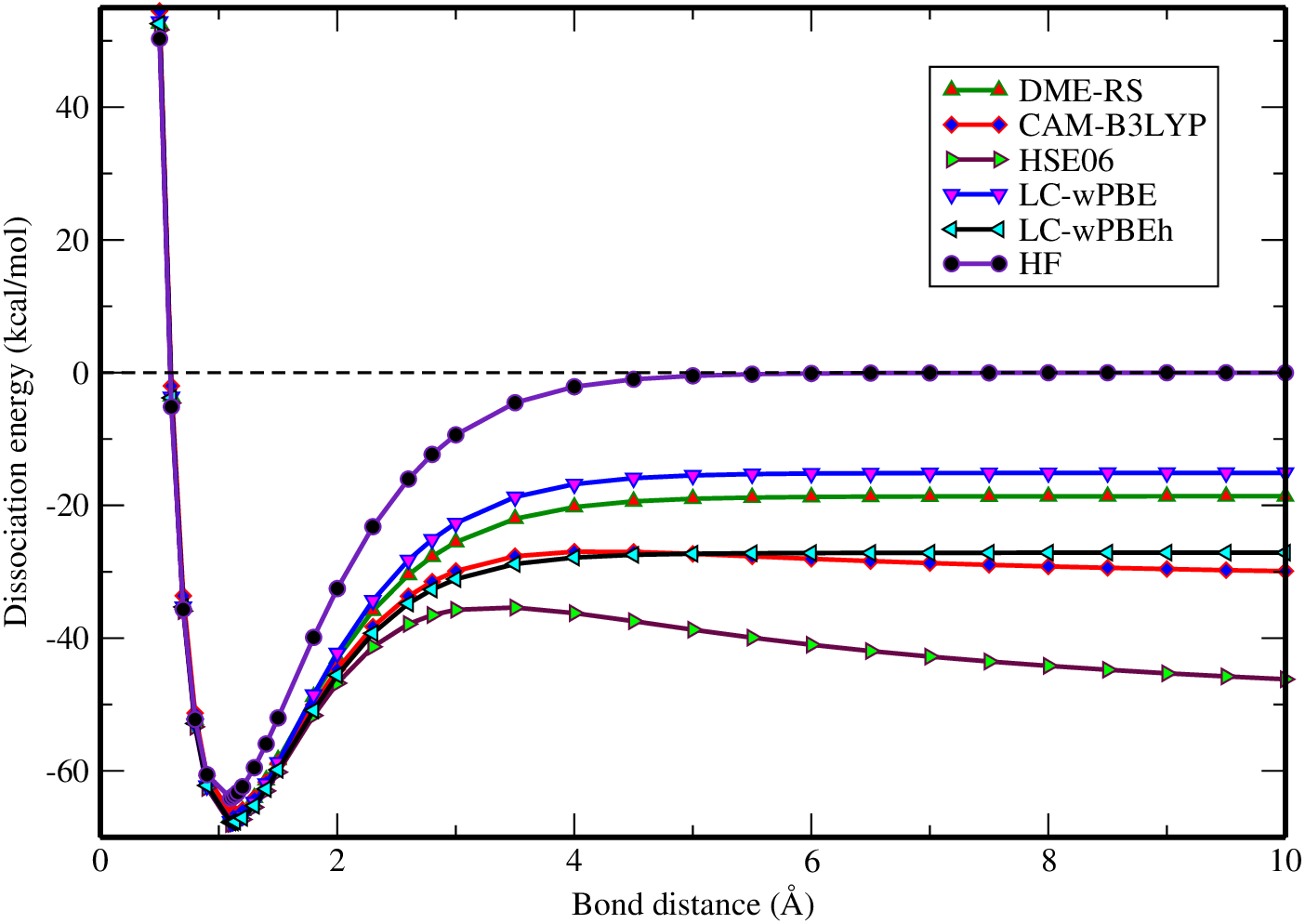}
%    \caption{Flower one.}
  \end{minipage}
  %\hfill
  \vspace{3cm}
  \begin{minipage}[b]{0.4\textwidth}
    \includegraphics[width=\textwidth]{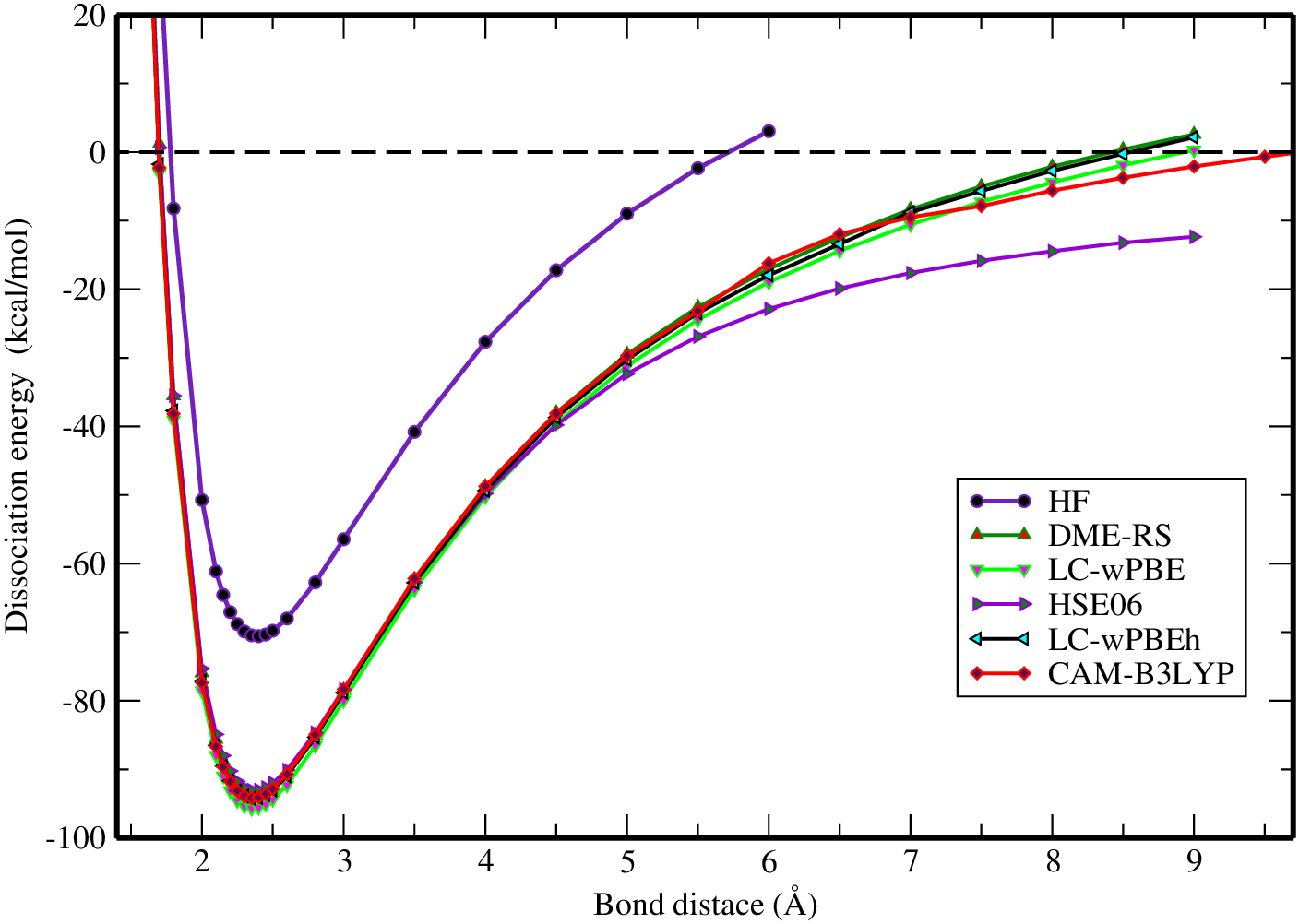}
    \caption{Dissociation curves of $H_2^+$ (upper panel) and  NaCl (lower panel) obtained using 6-311++G (3df,3pd) basis set. Zero level is set to be E(H) and E(Na)+E(Cl) 
respectively for both the cases.}
  \end{minipage}
\end{figure}

% \begin{figure*}%[h]
%  \centering
%\begin{subfigure}{.5\textwidth}
%  \centering
%  \includegraphics[width=0.90\linewidth]{diss.eps}
%  \caption{}
%  \label{fig:sub1}
%\end{subfigure}%
%\begin{subfigure}{.5\textwidth}
%  \centering
%  \includegraphics[width=.90\linewidth]{nacl.eps}
%  \caption{}
%  \label{fig:sub2}
%\end{subfigure}
%\caption{Dissociation curves of a) $H_2^+$ and b) NaCl obtained using 6-311++G (3df,3pd) basis set. Zero level is set to be E(H) and E(Na)+E(Cl) 
%respectively for both the cases.}
%\label{fig:test}
%\end{figure*}

\section{Conclusion}
In this paper, we have developed a new long range correction scheme for meta-gga exchange functionals in density-functional theory. 
The new scheme is based on Savin's long-range correction scheme using Tao-Mo exchange hole at the short range and Hartree-Fock 
exchange integral at the long range by separating electron-electron interaction operator $1/r_{ij}$ into short and long range 
by introducing standard error function. We compare our results with four other popularly known LC functionals like CAM-B3LYP, 
HSE06, LC-$\omega$PBE, LC-$\omega$PBEh. Among these functionals, CAM-B3LYP and LC-$\omega$PBEh mix both SR and LR Hartree-Fock exchange. In HSE06, 
a finite amount of HF exchange is used at SR but none in the LR limit, in order to scale down the computational cost of the 
non-local exchange integral for the extended system. Our DME-RS functional is of LC-$\omega$PBE type, where we have only one empirically 
fitted parameter $\mu$. On the other hand both CAM-B3LYP and LC-$\omega$PBEh has two extra parameters except $\mu$, which determine how 
much HF exchange is mixed at SR and LR. This functional gives comparable result with other heavily parametrized range separated 
functionals like $\omega B97X$~\cite{wb97x}(14 parameters), M11~\cite{m11} (40 parameters).

In solid state physics as the density tail is not that much important due to screened\cite{screening} coulomb interaction,
TM hole can be used at long range keeping HF treatment at short range\cite{HSE03}. This approach is also computationally 
advantageous for extended system applications.
\section{Supplementary material}
See supplementary material for result of the individual species for each cases.

\section{Acknowledgment}
The authors would like to acknowledge the financial support from the Department of Atomic Energy, Government of India.
%\pacs{}

\end{document}